\documentclass[twocolumn,showkeys,preprintnumbers,amsmath,amssymb]{revtex4}

\usepackage{graphicx}
\usepackage{dcolumn}
\usepackage{bm}

\begin{document}

\preprint{}

\title{Implicit Density Functional Theory}

\author{Bin Liu}
 \affiliation{Physics Department, New York University, 4 Washington Place, New York 10003}
\author{Jerome K. Percus}
\affiliation{Courant Institute of Mathematical Sciences, 251 Mercer St\\
and Physics Department, New York University, New York 10012
}
\date{\today}
\begin{abstract}
A fermion ground state energy functional is set up in terms of particle density, relative pair density, and kinetic energy tensor density. It satisfies a minimum principle if 
constrained by a complete set of compatibility conditions. A partial set, which thereby results in a lower bound energy under minimization, is obtained from the solution of model 
systems, as well as a small number of exact sum rules. Prototypical application is made to several one-dimensional spinless non-interacting models. The effectiveness of ``atomic" 
constraints on model ``molecules" is observed, as well as the structure of systems with only finitely many bound states. 
\end{abstract}


\maketitle
\section{Introduction}
The ground state properties of many-fermion systems, even with the implied restrictions on temperature domain, stationarity, homogeneity of particle type, have a physical 
importance that can hardly be overestimated. This has spawned a large variety of approximative, and in principle exact analytical and numerical solution techniques. There are also, 
in practice, conceptual restrictions on such techniques. They should not supply appreciably more information (which costs more as well) than one can conceivably use, but the 
intermediate constructs employed should be sufficiently nuanced that qualitative distinction between distinct systems can emerge, and ``physical intuition" utilized and developed 
along the way. 

One ``traditional" basically analytical direction that has been, and is being, pursued intensively is based upon the Rayleigh - Ritz variational principle. It is extreme in both 
aspects alluded to above: the intermediate construct used is the full $N$-body trial wave function $\psi_N$, and the exactly bounded information obtained is the ground state energy 
$E_0$ alone:
\begin {equation}
E_0 \le \left<\psi_N\right|\hat{H}_N\left|\psi_N\right>;
\end{equation}
here, $H_N$ denotes the system Hamiltonian. To be sure, bounding principles for expectations are also available but are much more conservative, the ground state restriction is 
trivially weakened to include ground state conditioned by compatible constants of the motion, and highly accurate ground state energy engenders confidence that expectations using 
the computed wave function will be highly reliable. And of course, the generality of this approach makes application to multi-species systems quite direct, but also very 
computation-intensive.

As it has become necessary to deal with increasingly complex structures, the electronic component of a Born - Oppenheimer macromolecule being an outstanding example, density 
functional techniques have emerged as very powerful semi-analytic tools. They use as intermediate constructs, and produce as output, the same quantity, the fermion (electron) 
density
\begin{equation}\label {density1b}
\begin{split}
\rho_N(x)&=\left<\sum_{j=1}^N \delta(x_j-x)\right>\\
&=\left<\psi^\dagger (x) \psi (x)\right>,
\end{split}
\end{equation} 
in first and second quantized form respectively, $x$ denoting all degrees of freedom of a particle. More elaborate versions work instead with the reduced one - body density matrix
\begin{equation}
\left(x\right|\gamma_N\left|x'\right)= \left<\psi^\dagger (x) \psi (x')\right>, 
\end{equation}
and attend to the spin population as well, but all depend upon intelligent semi-empirical information as to the dependence of the energy on this quantity, corresponding to a 
bounding principle of not much more than pictorial significance. In practice, even primitive local density functionals often give tolerable results, and density gradient extensions 
\cite{1} can, in professional hands, be remarkably effective, as well as be generalizable to to non-ground states and dynamical information.
 
An elusive goal for many years has been the establishment of practical analytical techniques which work through not much more than the needed output information and yet are also 
exact bounding principles at some level. The modifier ``practical" is where the difficulty lies. The version that has been pursued even longer than density functional techniques is 
that involving only the ground state 2-body reduced density matrix
\begin{equation}
\left(x_1, x_2,\right|\Gamma_N \left|x_1', x_2'\right)=\left<\psi^\dagger(x_2)\psi^\dagger(x_1)\psi(x_1')\psi(x_2')\right> 
\end{equation} 
of a pair-interacting system of the form
\begin{equation}
\begin{split}
\hat{H}_N&=\hat{h}_1 + \hat{h}_2,\\
\hat{h}_1&= \sum_{i=1}^N ¡¢\left[T(\hat{p}_i)+v(\hat{x}_i)\right]\\
\hat{h}_2&=\sum_{1\le i < j \le N} u(\hat{x}_i, \hat{x}_j).\\
\end{split}
\end{equation}
Then one has, in obvious notation, 
\begin{equation}\label{energymatrix}
\begin{split}
E_0=\min_{\gamma_N, \Gamma_N} \textrm{Tr} (K \gamma_N) + \textrm{Tr}(u\Gamma_N),\\
\mbox{where $K=T+v$ .} 
\end{split}
\end{equation}
But the reduction of an $N$-body system to the detailed properties of an effective $2$-body system involves a tremendous reduction of information, which must therefore be supplied 
indirectly \cite{16}. In the version referred to, these take the form of a small selection of known sum rules, and a large implicit, but only partially known, selection of 
inequalities that the 2 body density matrix must satisfy. It is only with the advent of modern computational tools - hardware and algorithms - that this approach has become 
feasible, but still primarily in the area of small systems and coarsely discretized function space.
 
In the present paper, we study an offshoot of this last activity, initially restricted to the very large domain of systems that can be modeled by having Newtonian kinetic energy
\begin{equation}
T(p)=p^2/2m
\end{equation}
as the only momentum - dependent contribution. The intermediate quantities we deal with are various one-point densities, particle density (\ref{density1b}) as well as suitably 
defined kinetic energy tensor density and pair interacting density (The pair distribution integrated over center of mass). These are minimally informative, but sufficient to 
represent major quantities of physical interest - including the energy, thereby allowing an energy bounding principle to be formulated. The reduction in information is even more 
extreme than in the pair-density matrix formulation, and so even more subsidiary restrictions must be imposed. But we will see that they can be selected by attention to the 
underlying physics, and then are rewardingly effective. This means that we will take great advantage of simple physical systems that share physical characteristics with the system 
at hand. An important tool will stem from the observation that any solvable model -- each of which is associated with an inequality restriction -- expands to a whole class of 
models under coordinate transformations.

In our current entree to this approach, we will confine our attention to very primitive ``toy" models, selected to probe the effectiveness of the technique without undue 
complexity. Thus, starting in Sec II, we deal only with one-dimensional non-interacting spinless fermions, establishing the basic sum rules and the model-generated form of 
inequalities. As will be reported in work now in progress, extension to 3 dimensions and spin is quite direct, if a bit more complicted, whereas the explicit inclusion of 
interaction requires a simple but non-trivial expansion of the technique (not even needed if only the construction of the Kohn-Sham pseudo-one-body density matrix \cite{6} is at 
issue). In particular, the required amassing of model reference systems is more intricate, and will be reported on in a later publication; see however Section V. In Sec III, we 
apply the technique to a few very elementary examples, and indicate in Sec IV how the strict minimization can be relaxed. Sec V tests to what extent toy ``molecules" can be solved 
via knowledge of their component ``atoms", and in Sec VI, we study the much more demanding situation in which only a finite number of non-interacting bound state exists.\\

\section{Implicit density functional inequality taken as constraint}
Since one knows, e.g. from the work of Hohenberg and Kohn \cite{7}, that one can use the $N$-electron density function $\rho_N(x)$ as the only variable when interaction and 
external potentials $u$ and $v$ are fixed, let us rewrite Eq.(\ref{energymatrix}) as 
\begin{equation}\label{density functional}
E_0=\min_{\rho_N}\left\{ T[\rho_N]+ \int{\rho_N(x)v(x)}dx + U_{\mbox{int}}[\rho_N]\right\}, 
\end{equation}
where $T[\rho_N]$ is the kinetic energy part and $U_{\mbox{int}}[\rho_N]$ the interacting part of the energy expectation. The in-principle separation into 
$T[\rho_N]+U_{\mbox{int}}[\rho_N]$ is a consequence of the fact that $\rho_N$ determines $v$ and hence the full wave function $\psi_N$ as well; $T[\rho_N]$ is not identical to the 
adiabatically interactionless kinetic energy relevant to Kohn-Sham. 

Therefore, the ground state can be obtained by applying a variational principle to the expected value of energy with respect to the density function $\rho_N(x)$, provided that the 
full functional of density is known. In most cases, the exact form of the functional is impossible to write down explicitly. And as we will emphasize in this paper, it's also 
unnecessary to do so since the functional is implied by sufficient many equalities or inequalities, several of which we apply as constraints on the minimization. 

\begin{figure}[hbp]
\begin{center}
\includegraphics[width=0.45\textwidth]{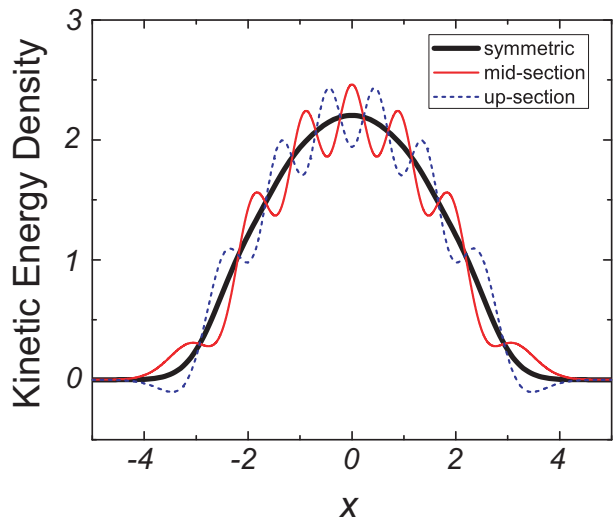}
\caption{(Color online)Three types of kinetic energy density for the ground state of harmonic system with 6 non-interacting fermions are shown here. The thickest dark line 
represents the 
symmetrized kinetic energy density, while the other two are the up- and mid-section of the symmetric kinetic energy density (Eq.(\ref{ts})). \label{fig:ked}}
\end{center}
\end{figure}

Consider a one dimensional non-interacting $N$-fermion reference system on coordinate space $\{X\}$, the ground state energy $E_N$ of which is known and must satisfy
\begin{equation}
\left<\psi_N\right|\hat{H}_N\left|\psi_N\right>\ge E_N,
\end{equation}
where $\hat{H}_N=\sum_{i=1}^N (\frac{1}{2}\hat{P}_i^2+V(\hat{X}_i))$, and $\psi_N$ is any anti-symmetric $N$-body wave function. An enormous convenience is that each solved 
reference generates a whole class of useful references, as follows: A continuous transformation of coordinates $X_i=f(x_i)$ can always be extended to a unitary 
tranformation\cite{13} by setting
\begin{equation}
\begin{split}
P_i&=\frac{g(x_i)p_i+p_ig(x_i)}{2},\\
\end{split}
\end{equation}
where $g(x)=1/J(x)$ and $J(x)=f'(x)$ is the transformation Jacobian. Hence the commutator relation $\left[x_i, p_j\right]=i\hbar\delta_{ij}$ persists, and a new set of canonical 
coordinates $\{x_i, p_i\}$ is obtained. Define the symmetric kinetic energy density operator(which becomes a tensor density in higher dimensional space) as 
\begin{eqnarray}\label{ts}
{\hat{T}(X)}=&&\frac{1}{8}\sum_{i=1}^N\left(\delta(X-\hat{X_i})\hat{P}_i\hat{P}_i+2\hat{P}_i\delta(X-\hat{X_i})\hat{P}_i\right.\nonumber\\
&&+\left.\hat{P}_i\hat{P}_i\delta(X-\hat{X_i})\right).
\end{eqnarray}
Eq.(\ref{ts}) is not a unique representation of kinetic energy density, since we can add any spatial divergence to it and yield the same total energy. For example, one has 
equivalent symmetrized up-section 
\begin{equation}
\hat{T}_u(X)=\frac{1}{4}\sum_{i=1}^N \left[\delta(X-\hat{X_i}),\hat{P}_i\hat{P}_i\right]_{\textrm{\tiny{+}}},
\end{equation}
or mid-section
\begin{equation}
\hat{T}_m(X)=\frac{1}{2}\sum_{i=1}^N \hat{P}_i\delta(X-\hat{X_i})\hat{P}_i.
\end{equation}
Both lead to the same total kinetic energy but the kinetic energy density is different. For a non-interacting fermion system with harmonic oscillator external potential well, the 
three types of kinetic energy densities mentioned above are shown in Figure~\ref{fig:ked}. Compared to the other two, the shell details of the fully symmetrized kinetic energy 
density are diminished. The advantage of picking up the symmetric one is that it gives a relative simpler form of the coordinate transform for kinetic energy.

Applying the symmetrized kinetic energy density definition (\ref{ts}), we have for the energy expectation for the previous Hamiltonian 
\begin{equation}
\begin{split}
\left<\psi_N\right|\hat{H}_N\left|\psi_N\right>&=\int T(X)dX + \int \rho_V(X) V(X)dX\\
											   &=\int g^2(x)t(x)dx+\int\rho_v(x)v(x)dx\\
                                               &+\frac{1}{8}\int\rho_v(x)g'^2(x)dx, \\
\end{split}
\end{equation}
where $T$ and $\rho_V$ are kinetic energy and electron density for the exactly solved reference system, $t(x)$ is the transformed kinetic energy density obtained with the symmetric 
kinetic energy density operator, defined the same way as $\hat{T}(x)$:
\begin{eqnarray}
\hat{t}(x)=&&\frac{1}{8}\sum_{i=1}^N\left(\delta(x-\hat{x_i})\hat{p}_i\hat{p}_i+2\hat{p}_i \delta(x-\hat{x_i})\hat{p}_i\right.\nonumber\\
&&+\left.\hat{p}_i\hat{p}_i\delta(x-\hat{x_i})\right),
\end{eqnarray} \\
Moreover, $v(x)=V(X)$ and $\rho_v$ is transformed from the original density function as $\rho_v(x)=\rho_V(X)J(x)$.

Therefore, for any realizable, i.e. ``$N$-representable" combination of kinetic energy density $t(x)$ and density $\rho(x)$, we must have
\begin{equation} \label{transform01}
\begin{split}
\int g^2(x)t(x)dx+&\int \rho(x)V\left(f(x)\right)\\
+&\frac{1}{8}\int \rho(x)g'^2(x) dx\ge E_N[V] ,
\end{split}\end{equation}
where $f'(x)=J(x)=1/g(x)$. Since the transform $J(x)$ is arbitrary, we have an inequality to be satisfied by any $N$-representable combination of $t(x)$ and $\rho(x)$:
\begin{equation}\label{constraint01}
\begin{split}
\min_g \int g^2(x)t(x)dx+&\int \rho(x)V\left(f(x)\right) dx\\
+&\frac{1}{8}\int \rho(x)g'^2(x) dx \ge E_N[V] ,
\end{split}
\end{equation}\\
With $V(X)$ and $E_N[V]$ given by the reference system, we have a well defined constraint for the functional relation between $t(x)$ and $\rho(x)$. A lower energy bound is obtained 
by carrying out the minimum with only the constraint satisfied, because we certainly have not included all constraints needed to guarantee $N$-representability. 

For a given v-representable (belonging to the potential v) density function $\rho_v(x)$, according to inequality (\ref{transform01}), the coordinate transform leads to

\begin{equation}\label{FV}
\begin{split}
F[V]&\equiv\int {g_V}^2(x)t(x)dx+\frac{1}{8}\int \rho(x){{g_V}'}^2(x) dx - E_N[V]\\
&\ge -\int \rho_v(x)v(x) dx ,
\end{split}
\end{equation}
where $g_V(x)=V'\left(V^{-1}\left(v\left(x\right)\right)\right)/v'(x)$. As this inequality is ubiquitous for any $V(X)$, we must have  
\begin{equation}\label{FVMIN}
\min_V F[V]+\int \rho_v(x)v(x) dx\ge 0. 
\end{equation}
According to Eq.(\ref{FV}), it's obvious that the minimum for $F[V]$ will be achieved if $g(x)=1$ where the inequality Eq.(\ref{FVMIN}) becomes an equality. Approaching its 
minimum, we have $\frac{\delta F[V]}{\delta V(x)}=0$, which gives
\begin{equation}
\begin{split}
\frac{\partial}{\partial x}&\left[g^2(x)\left(g(x)t(x)-\frac{[\rho'(x)g'(x)]'}{8}\right)\right]\\
                           &=-\frac{1}{2}\rho_v(x) v'(y)g(x)
\end{split}
\end{equation}
According to the statement above, letting $g(x)=1$, we instantly have
\begin{equation}\label{HyperVirial}
t'(x)=-\frac{1}{2}\rho_v(x) v'(x)
\end{equation}
Integrating on both sides, then
\begin{equation}\label{Virial_Int}
\begin{split}
\int t(x)dx=&-\frac{1}{2}\int_{-\infty}^{\infty}dx\int_{-\infty}^x dx'\rho_v(x')v'(x')\\
           =&-\frac{1}{2}\int_{-\infty}^\infty dx'\int_{x'}^\infty dx \rho_v(x')v(x')\\
	       =& \frac{1}{2}\int_{-\infty}^\infty x'\rho_v(x')v'(x')dx', 
\end{split}
\end{equation}
which coincides with the virial theorem \cite{8}. Stronger than the virial theorem, Eq.(\ref{HyperVirial}) provides another functional relation (see e.g. Baltin \cite{14}, March 
and Young \cite{15} to be satisfied by the pair of kinetic energy density and density so that they can possibly be v-representable. \\

\section{Primitive Applications}
Due to the last nonlinear term on the left-hand-side in the constraint Eq.(\ref{constraint01}), it can not easily be simplified. As to its positivity, by making the constraint a 
bit stronger, we can eliminate this term so that the feasibility of the constraint can be illuminated(Empirically, the correction due to the $(g')^2$ term always turns out to be 
very small - for an exception, see Eq.(\ref{asymcl})). Minimizing the left-hand-side with respect to the transform $g(x)$, we have the constraint on $t(x)$ and $\rho(x)$ simplified 
as 
\begin{equation}\label{constraint01s}
R_V[\rho, t]\equiv \int \left[\frac{\varphi^2t}{4}\right]^{1/3}dx+\int \rho V\left(f\right) dx \ge E_N[V], 
\end{equation}
where $\varphi$ and $f$ are functions of $x$, and
\begin{eqnarray}\label{constraint01sc}
\varphi'&=&-\rho V'(f) \nonumber \\ 
f'&=&\left(\frac{2t}{\varphi}\right)^{1/3} . 
\end{eqnarray}

The ordinary differential equation(ODE) array above won't challenge numerical calculations at all. However, in order to present the constraint Eq.(\ref{constraint01s}) concretely, 
let's consider a reference system with a linear external potential $V(x)$ in the one dimensional half space. That is 
\begin{equation}\nonumber
V(X)=\left\{\begin{array}{ll}
				X & \textrm{if $X\ge 0$,}\\
				\infty & \textrm{otherwise}	
				\end{array}
				\right.
\end{equation}
With this, the ODE array (Eq.\ref{constraint01sc}) is instantly solved, and the constraint can be further simplified to 
\begin{equation}
R_L[\rho, t]\equiv \frac{3}{2} \int_0^\infty{(A(x))^{2/3}(2t(x))^{1/3}}dx \ge E_L(N),
\end{equation}
where $A(x)$ is the cumulative density, defined as 
\begin{equation}
A(x)\equiv\int_x^\infty \rho(x')dx', 
\end{equation}
$E_L(N)=-\sum_{i=1}^N 2^{-1/3}a_i$, the ground state energy for N non-interacting fermions with a linear well as external potential, $a_i$ is the ith root of the first kind of Airy 
function \cite{9}. As the number of fermions $N$ increases, asymptotically, we have $E_L(N)\approx \frac{3}{10} (3\pi)^{2/3}N^{5/3}$, with extremely rapid convergence. 

If we are interested in the asymptotic value of the ground state energy, we can apply this to finding the ground state energy of a non-interacting fermion system with half space 
external potential $v(x)=\frac{1}{\gamma}x^{\gamma}$ 
\begin{equation}\label{41}
\min_{\rho_N}E[\rho_N]=\min_{\rho_N}\frac{\gamma+2}{2\gamma}\int{\rho_N(x)x^\gamma}dx
\end{equation}
with the constraint
\begin{equation}\label{42}
R_L[\rho, t]=\frac{3}{2} \int{A^{2/3}(2t)^{1/3}}dx \ge \frac{3}{10} (3\pi)^{2/3}N^{5/3}, 
\end{equation}
where the local virial constraint Eq.(\ref{HyperVirial}) has also been applied. 
It can be shown that the minimum of Eq.(\ref{41}) occurs only at the boundary of the function space of the density $\rho_N(x)$. And since the density function is non-negative, we 
have 
$\rho_N(x)=N\delta(x-x_0)$. From Eq.(\ref{42}), we have
\begin{equation}\label{43}
x_0\ge \left(\frac{1}{5}\right)^{\frac{3}{\gamma+2}}(3\pi N)^{\frac{2}{\gamma+2}}
\end{equation}
Therefore,
\begin{equation}\label{EN}
E[\rho_N]/N\ge \frac{\gamma+2}{2\gamma} \left(\frac{1}{5}\right)^\frac{3\gamma}{\gamma+2}(3\pi N)^\frac{2\gamma}{\gamma+2}
\end{equation}
Not surprisingly, when $\gamma=1$, we have 
\begin{equation}
E[\rho_N]/N\ge \frac{3}{10} (3\pi N)^{2/3}, 
\end{equation}
which is identical with the exact asymptotic behavior of the system with the half space linear well as external potential. For the Coulomb potential, where $\gamma=-1$, 
Eq.(\ref{EN}) asserts that
\begin{equation}\label{asymcl}
E_C[\rho_N]/N\ge - \frac{125}{18 \pi^2}N^{-2},
\end{equation}
which is clearly false because $E_C[\rho_N] \rightarrow$ - const as $N \rightarrow \infty$. The reason for this apparent paradox lies in the neglected $(g')^2$ term, which is 
ordinarily very small, but when it is required to map a density due to a regular potential onto one from a singular potential, this is no longer the case, and the inclusion of 
$(g')^2$ could be mandatory. We have taken the $\gamma=-1$ case as a first example to show that the reference system must be reasonably similar to the system under study to make 
sense. Ignoring the $(g')^2$ contribution accentuates this difference to the point that a lower bound is no longer obtained, but the general comment remains solid. We now consider 
further examples that develop this implicit criterion. 

For harmonic oscillating fermions with $\gamma=2$, we have
\begin{equation}
E_{HO}[\rho_N]/N\ge \left(\frac{1}{5}\right)^{3/2} {3\pi}N \approx 0.842978 N
\end{equation}
compared to the exact $1.0 N$.

As $\gamma \rightarrow \infty$, we have the rigid wall box to solve, the constraint gives
\begin{equation}
E_{RB}[\rho_N]\ge \frac{9\pi^2}{250} N^3
\end{equation}
compared to the exact asymptotic $E_{\mbox{rigid box}}=\frac{\pi^2}{6}N^3$. 
As we see, with one constraint from the linear well reference system alone, applying the minimization scheme won't result in a bound very close to the exact solution.\\
If we subject the density minimizing the energy functional above to a different constraint, for example taking harmonic oscillating fermions as a reference system, we have 
\begin{equation}
\begin{split}
R_{HO}[\rho,t]=&\min_{f(x)}\left(\int_0^{x_0} \frac{N}{2(f'(x))^2}dx {x_0}^{\gamma-1}\right.\\
&\quad+\left.\frac{N}{2} (f(x_0))^2\right)\\
=&N {x_0}^{\frac{\gamma+2}{2}}
\end{split}
\end{equation}
Substituting $x_0$ from Eq.\ref{43}, then
\begin{equation}
R_{HO}[\rho, t]=N\left(\frac{1}{5}\right)^{3/2}(3\pi N)\approx 0.84 N^2
\end{equation}  
Thus, for the linear constraint optimum, constraint $R_{HO}[\rho,t] \ge N^2 $ is badly unsatisfied. As will be shown later, the ground energy level obtained with the minimizing 
scheme will be dramatically improved if more reference systems are involved. \\

\section{Density Configuration Concentration}
For any $N$-representable density profile $\rho(x)$, there always exist scale transformed density functions $\rho_\alpha (x)\equiv \alpha \rho(\alpha x)$ so that using virial 
constraint (\ref{Virial_Int}) the applied reference dependent constraint is satisfied. Among these transformed densities, we have a uniquely defined reference dependent density 
functional
\begin{equation}
\begin{split}
E_{\mbox{Ref}}[\rho]\equiv& \min_{\tiny{\begin{array}{c}
\alpha\mbox{ satisfying}\\
\mbox{the constraint}\\
\end{array}}
} \left(\int \mbox{KED}_{\mbox{HV}}\left(x;[\rho_\alpha]\right)dx\right.\\
&\quad+\left.\int{\rho_\alpha(x)v(x)}dx \right), 
\end{split}
\end{equation}
where $\mbox{KED}_{\mbox{HV}}$ is the kinetic energy density term with the local hyper-virial constraint (\ref{HyperVirial}) applied. Therefore, a pseudo energy landscape (PEL) of 
the density profile is obtained, the global minimum of which provides a lower bound to the energy expectation of the ground state. It is noticed that because of making use of the 
scale transform, the landscape $E_{\mbox{Ref}}[\rho]$ depends only on the scale free configuration of the density profile. Rather than searching the minimum over the entire 
non-negative density function space, one can obtain the identical minimum by searching within a more compact subspace provided that every configuration of density has been 
included. Therefore, the reference dependent landscape is furthermore reduced to the density configuration hypersurface. The minimization task can be achieved by searching the 
minimum over the randomly generated $n$ dimensional discrete density profile subspace (100000 configurations were typically used ) as 
\begin{equation}
\rho_0 \otimes \rho_1 \otimes \cdots\otimes \rho_{n-1} \sim 0 \otimes \underbrace{Rand(m)\otimes \cdots \otimes Rand(m)}_{n-2}\otimes 0  
\end{equation}
where both $n$ and $m$ are positive integers and $Rand(m)$ randomly generates an integer between $0$ to $m-1$. $\rho_0$ and $\rho_{n-1}$ are zero since the density vanishes at both 
ends in one dimensional space. The entire configuration hypersurface is covered as $n$ and $m$ approach infinity. To avoid the unnecessary computing effort in searching within the 
unsmooth function space, a smooth density function subspace can be obtained by interpolating the randomly generated control points with splines. Therefore, for practical use, 
finite number of random numbers is required to generate a density profile. In this paper, for random generated density profiles, a B-spline \cite{10} of degree 3 is applied.
 
\begin{figure*}
\begin{center}
\includegraphics[width=0.8\textwidth]{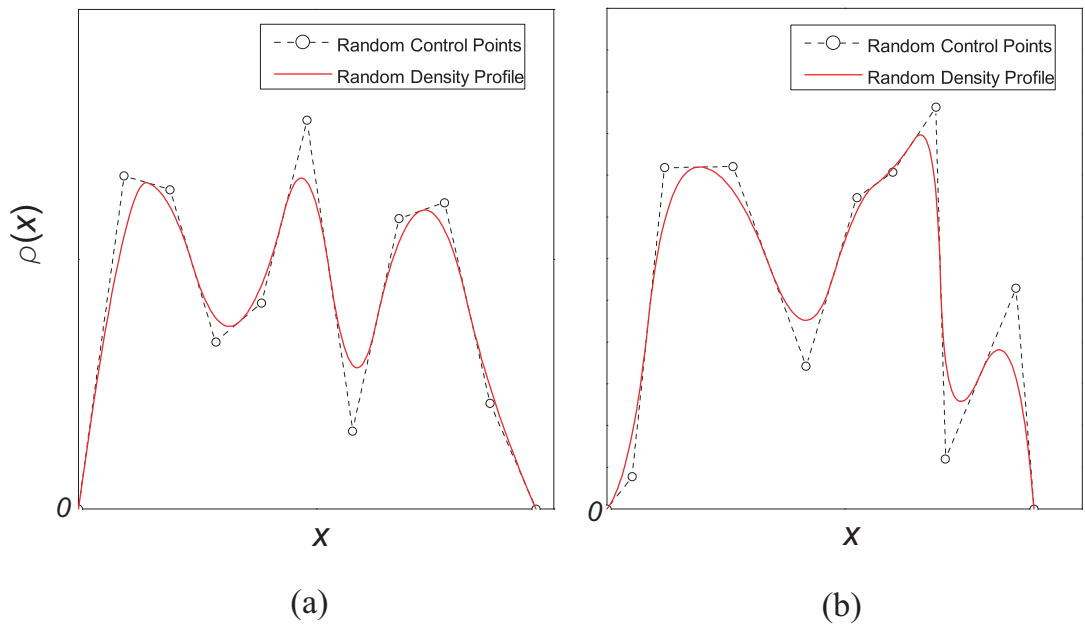}
\caption{(Color online)To illustrate how the random density configuration is generated, the solid lines show the random density profiles, which are provided by B-splines along the 
arbitrary selected set of randomly generated control points marked with `o's. (a) shows the instance with 11 control points evenly spaced while (b) shows the instance with 11 
control points 
randomly spaced. The density configuration space generated with evenly spaced control points is a subset of density configuration space generated with the same number of randomly 
spaced ones. \label{fig:randomdensity}}
\end{center}
\end{figure*}

As shown in Figure~\ref{fig:randomdensity}, the control points can be either equally spaced or randomly spaced to include more rapidly varying density configurations with the same 
number of control points. This is useful when, as in this introductory study, we want to economize on the resolution employed.

The lower bound of the ground state energy is given by 
\begin{equation}
E_L=\min_{\rho_1,\rho_2,\cdots \rho_{n-1}} E_{\mbox{Ref}}\left[\rho=\mbox{spline}\left(\rho_1,\rho_2,\cdots \rho_{n-1}\right)\right]
\end{equation}

\begin{figure}[hbp]
\begin{center}
\includegraphics[width=0.4\textwidth]{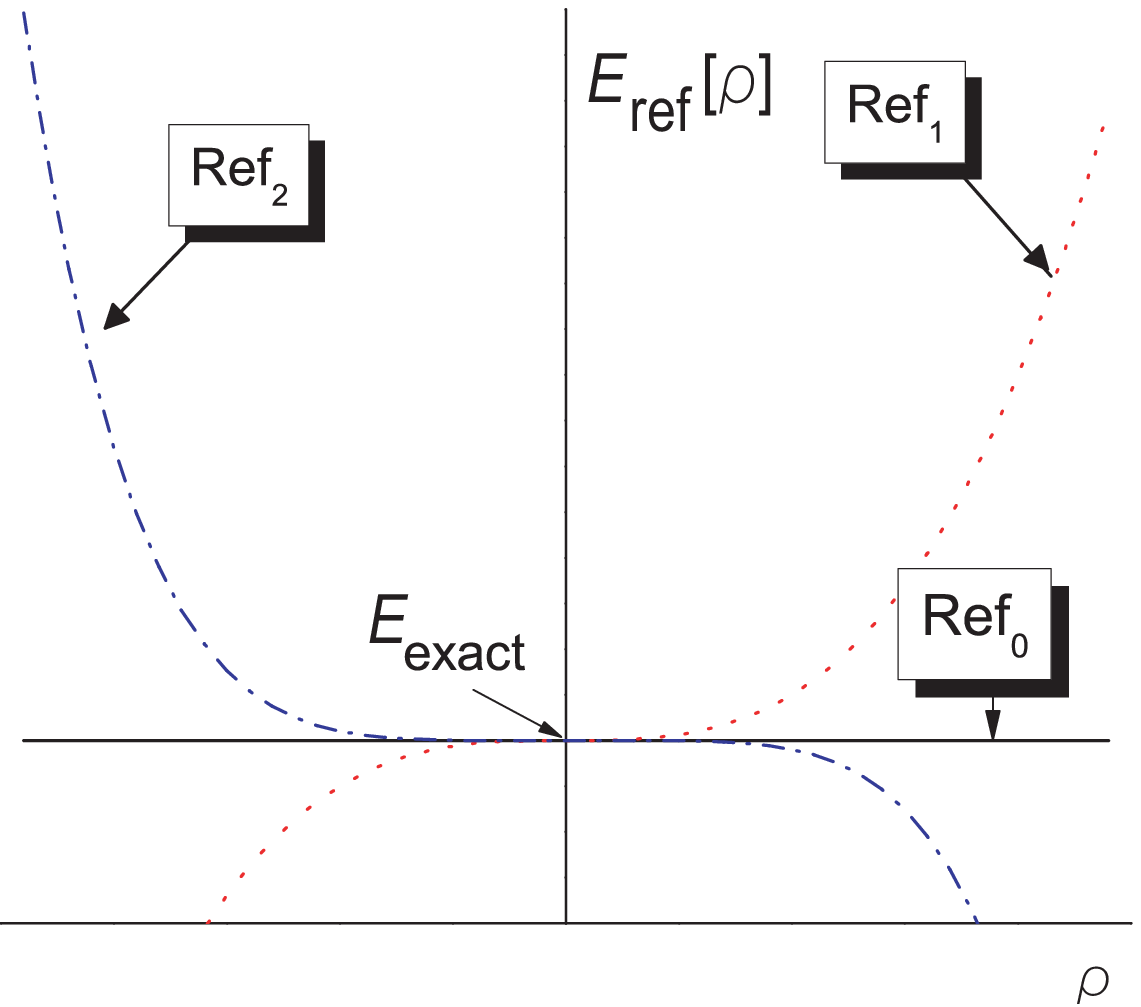}
\caption{(Color online)The cartoon shows the landscapes $E_{\mbox{Ref}}$ in density configuration space, which have dependence on different reference systems. For those reference 
systems with 
similar external potential to that of the target systems, such as $\mbox{Ref}_0$ shown in the figure, we have a flat landscape represented with a horizontal line. For reference 
systems very different from the target system being applied, the reference dependent landscapes are shown in the as dotted curves. All the landscapes concentrate at the exact 
ground state energy level of the target system.\label{fig:energylandscape}}
\end{center}
\end{figure}

According to the definition of the landscape $E_{\mbox{Ref}}$, if one were to take reference system with external potential identical to that of the target system, it's trivial 
that the reference dependent landscape $E_{\mbox{Ref}}[\rho]=E_{\mbox{exact}}$, where $E_{\mbox{exact}}$ is the ground state energy of the target system, which is flat all over the 
density configuration hypersurface. Therefore, any landscapes tangential to it have the landscape altitudes concentrating near the exact ground state energy, which is a saddle 
point. Note that for a reference system very different from the target system, there's only one crosspoint in the density configuration hypersurface between the reference dependent 
landscape and the flat landscape described above. It can be concluded that the distribution function of $E_{\mbox{Ref}}$ has its maximum at the ground state energy of the target 
system, provided none of the density configurations dominate when generating the density profiles randomly, as illuminated by Figure~\ref{fig:energylandscape}. Consequently, the 
randomly generated density configurations provides not only a lower bound but also practical approximation to the target system.

\begin{table}[hb]
\caption{Ground state energy of $N$-harmonic-oscillating-fermion obtained from reference system with linear well. $E_\textrm{exact}$ is the exact energy expectation solved for N 
non-interacting fermions, and $E_\textrm{most}$ is the energy expectation with highest concentration generated by random density profiles}\label{t1}
\begin{ruledtabular}
\begin{tabular}{ccc}
N & $E_\textrm{exact}\footnote{exact energy expectation}$ & $E_\textrm{most}$\footnote{most probable energy}\\
\hline
1 & 1.5 & 1.6\\
2 & 5 & 5.0\\
3 & 10.5 & 10.5\\
4 & 18 & 17.8\\
\end{tabular}
\end{ruledtabular}
\end{table}

Consider the non-interacting $N$-fermion system with $v(x)=\frac{x^2}{2}$ as the external potential to solve. Rather than taking the asymptotic constraint, one can make use of the 
exact energy level of N non-interacting fermions within the linear well. The pseudo energy landscape is generated with the $E_{\mbox{Ref}}$ associate with the random density 
profiles, which are B-spline interpolated. We find that the energy distribution concentrates at the following values shown by Table \ref{t1}. The exact value is provided as a 
comparison.

\section{Molecule-Like Systems }
In our primitive application of the energy minimizing scheme, we have shown that coordinate-transformations from a known reference system do provide a somewhat weak lower bound for 
the ground state energy level of a fairly different system, too. We are also interested in applying the constraint to a locally similar but globally different target system . Let's 
see what would happen if we convert a single well reference system into a molecule with double core.

For example, let's consider the fermion system to be solved as having the double linear well as external potential 
\begin{equation}\label{26}
v(x)=\left\{\begin{array}{ll}
				x-b & \textrm{if $x\ge b$,}\\
				b-x & \textrm{if $0\le x < b$}\\
				x+b & \textrm{if $-b\le x <0$}\\
				-b-x & \textrm{if $x<-b$}	
				\end{array}
				\right.
\end{equation}

The reference system has the external potential illustrated by Figure~\ref{fig:reference}. 
\begin{equation}\label{27}
V(x)=\left\{\begin{array}{ll}
				a(x-b) & \textrm{if $x\ge b$,}\\
				a(b-x) & \textrm{if $x < b$}
				\end{array}
				\right.
\end{equation}

\begin{figure}[h]
\begin{center}
\includegraphics[width=0.4\textwidth]{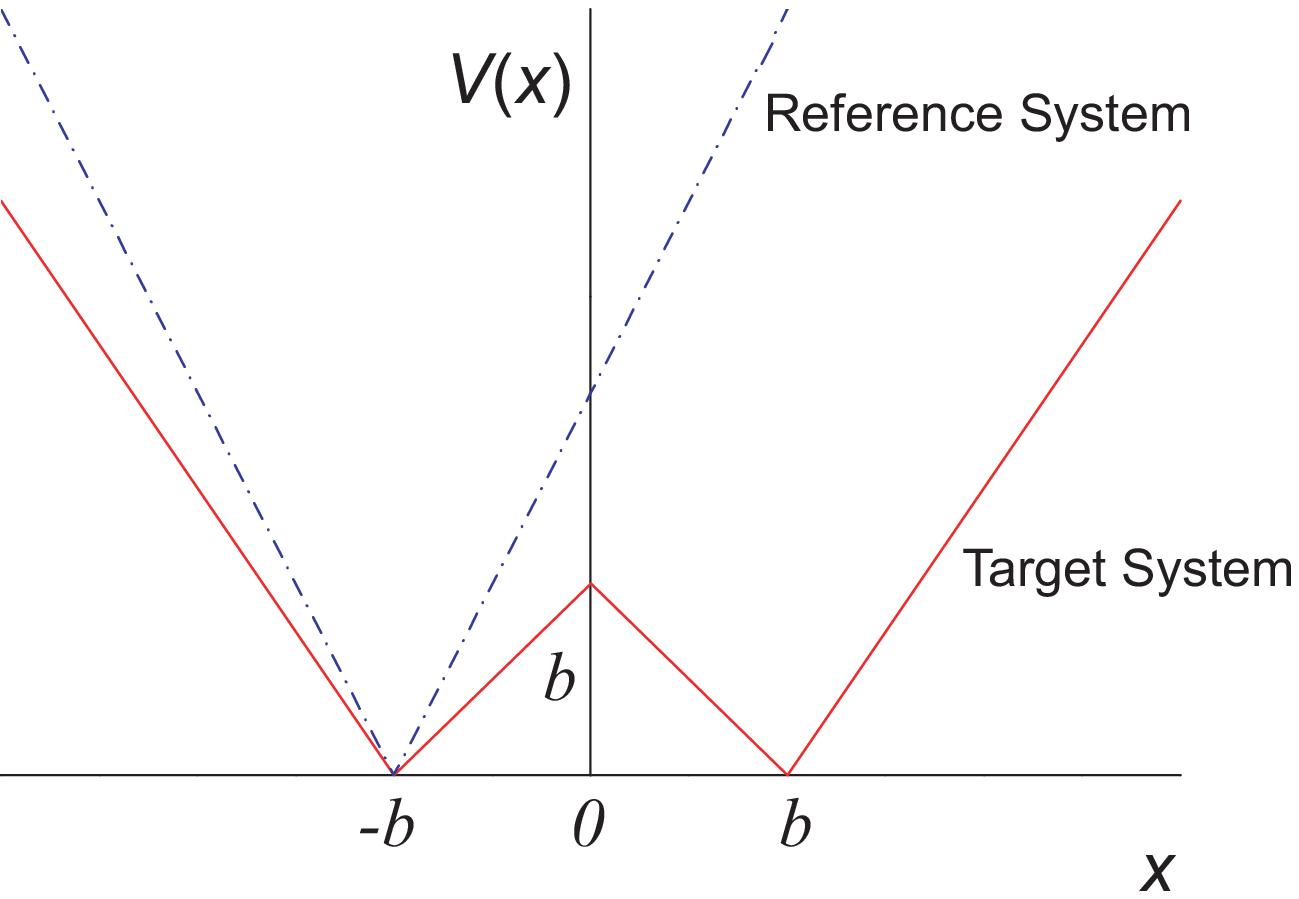}\label{fig:reference}
\caption{(Color online)An illustration of the potential of reference system and target system. \label{fig:reference}}
\end{center}
\end{figure}

By minimizing the energy subject to the single well constraint, the minimization take place at the boundary of the function space of density. 
\begin{equation}
\rho_N(x)=\frac{N}{2}\left(\delta(x-x_0)+\delta(x+x_0)\right),
\end{equation}
and $x_0$ is determined by the asymptotic constraint
\begin{equation}
\begin{split}
R_{LS}[\rho_N]&=\frac{3}{2}\int{(A_s(x))^{2/3}(2T(x))^{1/3}}dx\\
&\ge \frac{3}{10}\left(\frac{3\pi}{2}\right)^{2/3} N^{5/3}
\end{split}
\end{equation},
where 
\begin{equation}
A_s(x)=\left\{\begin{array}{ll}
				\int_x^\infty{\rho_N(x)}dx & \textrm{if $x\ge b$,}\\
				&\\
				\int_{-\infty}^x{\rho_N(x)}dx & \textrm{if $x < b$}
				\end{array}
				\right.
\end{equation}
We have
\begin{equation}
x_0\ge \frac{1}{5}\left(\frac{3\pi N}{2}\right)^{2/3}.
\end{equation}

And so we have for the ground state energy of the double linear well system
\begin{equation}
E[\rho_N]=\frac{3}{2}N x_0-N b\ge N\left(\frac{3}{10}\left(\frac{3\pi N}{2}\right)^{2/3}-b\right)
\end{equation}
as a lower bound. 

On observing the energy landscape concentration as shown before, we find the most probable energy levels, which can also be compared with the exact ground state energy level of the 
double linear well system. As shown in Table \ref{t2}, the most probable energy levels give the best approximation. With $N$ increasing, the minimization with single well 
constraint alone gives a closer result than the Thomas-Fermi approximation.

\begin{table}
\caption{Ground state energy of double linear well system with $b=1$ and single linear well reference. The second column is the exact energy for non-interacting fermions with 
double well as external potential. The third column is the energy at which the density configurations concentrate with a single well as reference system. The fourth column is the 
energy obtained by subjecting the minimization to the single well constraint, serving as lower bound. The fifth column provides the result given by Thomas-Fermi Approximation as a 
comparison}\label{t2}
\begin{ruledtabular}
\begin{tabular}{ccccc}
N & $E_\textrm{exact}\footnote{exact energy expectation}$ & $E_\textrm{most}$ single well\footnote{most probable energy with single-well constraint} & $E_\textrm{min}$ single well 
\footnote{lower bound with single-well constraint} & $E_\textrm{T-F}$ \footnote{energy expectation with Thomas-Fermi approximation}\\
\hline
1 & 0.6266 & 0.630 & -0.157 & 0.5312\\
2 & 1.6622 & 1.580 & 0.677	& 1.6118\\
3 & 3.5386 & 3.206 & 2.262 & 3.1791\\
4 & 5.9278 & 5.477 & 4.499 & 5.2861\\
5 & 8.9200 & 8.337 & 7.328 & 7.9197\\
6 & 12.421 & 11.74 & 10.71 & 11.059\\
7 & 16.430 & 15.67 & 14.60 & 14.684\\
8 & 20.916 & 20.08 & 18.98 & 18.775\\
9 & 25.861 & 24.95 & 23.84 & 23.317\\
10 & 31.252 & 30.28 & 29.14 & 28.295\\
11 & 37.069 & 36.05 & 34.88 & 33.696\\
12 & 43.305 & 42.23 & 41.04	& 39.508\\
13 & 49.944 & 48.82 & 47.61 & 45.720\\
14 & 56.979 & 55.81 & 54.57 & 52.323\\
15 & 64.399 & 63.19 & 61.93 & 59.309\\
16 & 72.196 & 70.94 & 69.67	& 66.668\\
\end{tabular}
\end{ruledtabular}
\end{table}

The potnetial consequence of the quite decent agreement at this preliminary level are far - reaching: the same strategy can be used for full interacting atoms in molecules, a 
situation in which universally useful model systems are few in number. 

\section{More Than Single Constraint And Shielded Coulomb Potential}
As stated above, since the density configuration for the system to be solved should agree with every reference system provided, the more constraints we apply, the better the result 
one would expect. Therefore, we would like to constrain the target system by two reference systems to see how the lower energy bound is improved. 

Among all the density functions which satisfy the constraints, we are interested in those that generate the lowest energy levels together with the assistance of the local 
hyper-virial theorem. These energy levels are used to provide a lower energy bound. The B-splines generated by random numbers are still used as the density configuration 
candidates. 

We will see that not all pairs of constraints are effective in providing a decent lower energy bound for the target system. Not surprisingly, only those pairs of constraints that 
bracket the target system are really effective.

Note that only the constraints from the reference system with external potential $v(x) \sim x$ can be written down explicitly, while the others we can make use of so far can only 
be applied numerically. A primitive test of the effect of the double constraint is by making use of the asymptotic behavior of the constraints when the number of particles is 
sufficiently large, and comparing the result with those obtained from Thomas-Fermi approximation. The latter is believed to be a good approximation to describe the energy levels 
for the non-interacting fermion system when the number of fermion is large enough, except for those with Coulomb potential as external potential. 

The target system we will test has external potential in the form $v(x)=\frac{x^\gamma}{\gamma}$. Pairs of constraints that can bracket the target system and those that can not are 
selected to generate the lower energy bounds for the target system. The results are shown in Table \ref{t3}. We see that the result doesn't improve much compared with those 
obtained from the single constraint unless the pair of references system can bracket this system. The role of ``bracketing" is overwhelmingly important.

\begin{table}
\caption{Lower energy bound obtained with constraints from two reference systems. The asterisk in the first column indicates that the pair of reference systems can bracket the 
target system. }\label{t3}
\begin{ruledtabular}
\begin{tabular}{cccccc}
$\gamma$ & $\textrm{Ref}^1$\footnote{the external potential of the applied reference system}  & $\textrm{Ref}^2$ &$E_\textrm{min}$\footnote{lower energy bound with the constraints 
applied} & $E_\textrm{most}$\footnote{most probable energy with double constraints}& $E_\textrm{T-F}$\footnote{asymptotic energy expectation with Thomas-Fermi Approximation}\\
\hline
$2^*$	& $x$ & $x^3$ & $ 0.99 N^2 $ & $1.03 N^2$ &$ 1.00 N^2 $\\
$2^*$	& $-x^{-1}$ & $x^3$ & $ 0.64 N^2 $ & $1.02 N^2$ &$ 1.00 N^2 $\\
$2$	& / & $x^3$ & $ 0.37 N^2 $ & / &$ 1.00 N^2 $\\
$2$	& $-x^{-1}$ & $x$ & $ 0.84 N^2 $ & $1.00 N^2$ &$ 1.00 N^2 $\\
$2$	& / & $x$ & $ 0.84 N^2 $ & / &$ 1.00 N^2 $\\
$1.5^*$	& $x$ & $x^2$ & $ 1.09 N^\frac{13}{7} $ & $1.11 N^\frac{13}{7}$ &$ 1.10 N^\frac{13}{7} $\\
$1.5$ & $x$ & / & $1.00 N^\frac{13}{7}$ & / & $1.10 N^\frac{13}{7}$\\
$3$ & $x$ & $x^2$ & $0.83 N^\frac{11}{5}$ & $ 0.93 N^\frac{11}{5}$ & $ 0.94 N^\frac{11}{5}$\\
$3$ & $x$ & / & $0.68 N^\frac{11}{5}$ & / & $0.94 N^\frac{11}{5}$
\end{tabular}
\end{ruledtabular}
\end{table}

As discussed above, we have shown that our continuous coordinate transform scheme works well with most fermion systems having an infinite number of bound states. It will be 
interesting to know if it will break down when the system has finitely many bound states. One such $N$-fermion system that was the focus of past study is the shielded Coulomb 
system\cite{11} with the external potential $v(x)=-\frac{e^{-r/D}}{r}$, which can only be numerically solved \cite{12}. It's known to have a finite number of bound energy levels 
due to the effect of screening, and the number of bound energy levels depends on the screening length $D$. 

The pair of reference systems we use here are Coulomb and harmonic oscillator systems. And the exact value of each energy level is applied for each reference system. For half space 
1-D Coulomb systems, the ground state energy level for $N$ fermions is proportional to $H(N,2)\equiv \sum_{n=1}^N \frac{1}{n^2} $, where $H(x, 2)$ is the harmonic number function 
of $x$ with order 2. And the ground state energy level for $N$ fermions within a half space 1-D harmonic well is proportional to $N(N+\frac{1}{2})$.

The results with random density configuration candidates are shown in Table \ref{t4}. We do have finite number of bound states with this pair of constraints, even though the 
reference systems that we apply have an infinite number of bound states. Only when the screening length arrives at some threshold length, can a new bound state survive from the 
double constraints.

\begin{table}[h]
\caption{Lower bound for the ground state energy with shielded Coulomb potential. The number to the left of each cell is the lower bound obtained with double constraints. The 
number to the right is the energy level obtained with numerical calculation.}\label{t4}
\begin{ruledtabular}
\begin{tabular}{cccccc}
& \multicolumn{5}{c}{$N$}\\
$D$ & $1$ & $2$ & $3$ &$4$& $5$\\
\hline
$2$ & $-0.244|0.148$ & /&/&/&/ \\
$5$ & $-0.397|0.327$ & $-0.419|0.339$&/&/&/\\
$10$& $-0.448|0.407$ & $-0.521|0.457$ & $-0.526|0.460$ &/&/\\
$20$& $-0.471|0.452$ & $-0.571|0.534$&$-0.602|0.553$ &$-0.607	|0.556$&/\\
\end{tabular}
\end{ruledtabular}
\end{table}
\section{Conclusion}
We have seen that the analysis of a many-fermion ground state can be recast as a constrained minimization of a functional of two scalar and one tensor field. For non-interacting 
fermions, only the density and a suitably defined kinetic energy tensor density are required, and the effort is shifted to tabulating and using the needed constraints, which can be 
tuned to emphasize known and hypothesized physical aspects of the system. Taking the one-dimensional spinless system as prototype, a class of inequalities based upon solvable 
models has been developed, as well as the hypervirial equality. With only a small number of solvable models to help us, quite decent results have been obtained for some simple 
systems, including mock ``molecular" ones and systems with a controlled number of bound states. Extension of these techniques to 3 dimensions, and to the spin degree of freedom, is 
direct, and will be reported in the near future. Extension to physical interaction is less direct, and several paths are under study, which will be reported as well. They include 
universal bounds on the interaction energy as a functional of $\rho$ and $t$ (see ref\cite{13} for a rudimentary example), and modification of the coordinate transformation tactic 
in the face of interactions. It must of course be emphasized that this continuing investigation is not to be regarded as competition to the array of high accuracy computational 
techniques that have been developed, but rather as a low-cost replacement when fine detail is not required.
\section{Acknowledgment}
The contribution of JKP was supported in part by DOE under grant DE-FG02-02ER15292.                      

\end{document}